\documentclass[prl,amsmath,amssymb,twocolumn,superscriptaddress,showpacs,nobalancelastpage]{revtex4-1}
\pdfoutput=1
\usepackage{amsmath,amssymb}
\usepackage[usenames]{color}
\usepackage{amssymb}
\usepackage{grffile}
\usepackage[pdftex]{graphicx}
\usepackage{amsmath, amstext, amssymb, amsfonts, amsxtra}
\usepackage{textcomp}
\usepackage{xspace}

\begin{document}

\title{Hidden string  order in a  hole-superconductor with extended correlated hopping}

\author{ Ravindra W. Chhajlany} 
\address{ICFO-Institut de Ci\`encies Fot\`oniques, Av. Carl Friedrich
Gauss 3, 08860 Barcelona, Spain}
\address{ Faculty of Physics, Adam Mickiewicz University, Umultowska 85, 
61-614 Pozna{\'n}, Poland}

\author{ Przemys{\l}aw R. Grzybowski}
\address{ Faculty of Physics, Adam Mickiewicz University, Umultowska 85, 
61-614 Pozna{\'n}, Poland}
\author{Julia Stasi{\'n}ska}

\address{ICFO-Institut de Ci\`encies Fot\`oniques, Av. Carl Friedrich
Gauss 3, 08860 Barcelona, Spain}
\author{Maciej Lewenstein}

\address{ICFO-Institut de Ci\`encies Fot\`oniques, Av. Carl Friedrich
Gauss 3, 08860 Barcelona, Spain}

\address{ICREA-Instituci\'o Catalana de Recerca i Estudis Avan\c cats, Lluis
Campanys 23,
08010 Barcelona, Spain}
\author{Omjyoti Dutta}
\address{Instytut Fizyki im. M. Smoluchowskiego, Uniwersytet Jagiello{\'n}ski, {\L}ojasiewicza 11, 30-348 Krak{\'o}w, Poland}

\begin{abstract}
Ultracold fermions in a one-dimensional, spin-dependent, optical lattice are described by a  non-standard Hubbard model with  next-nearest-neighbour correlated hopping. Periodic driving of the lattice allows wide tuning of the system parameters. We solve this  model exactly for a special value of the correlated  hopping. The solution reveals the general properties of this system for arbitrary filling:  exact and asymmetric spin-charge separation, a gapless spectrum of lowest energy excitations, a spin-gap, which may be interpreted in terms of collective hole pairing and a non-vanishing den Nijs-Rommelse type string correlator. Numerical studies away from the integrable point show the persistence of both long range string order and  spin-gap. 
\end{abstract}

\maketitle

\paragraph{Introduction -- } 
The identification and description of interesting collective states of many particles is the essence of  contemporary many body physics. 
The study of one-dimensional (1-d) systems holds  a special place in this effort \cite{Giamarchi-book}, where effects of enhanced quantum fluctuations are particularly important. 
These systems are amenable to various exact or quasi-exact analytical and precise numerical methods which  reveal  surprises  and exotic physics, 
 {\it e.g.} Luttinger liquids  and associated spin-charge separation \cite{Tomonaga,Luttinger,Haldane,Ogata}, fractional excitations \cite{Faddeev}, non-local symmetry breaking captured by string order parameters \cite{AKLT,Nijs}, etc. 
Due to recent progress in ultra-cold quantum gas  experiments, simulations of various theoretical models with unprecedented control are becoming possible. In particular, non-standard extensions of the paradigmatic Hubbard models  are highly desired  \cite{Dutta-review}. A notable example concerns correlated hopping  (CH) considered already in the context of multi-orbital physics \cite{Jurgensen}, dipolar interactions \cite{Sowinski} and time-periodic modulation of on-site interactions \cite{Santos}.

In this Letter, we investigate a  1-d fermionic model with an extended CH spanning next-nearest nieghbours. This model \textit{precisely} describes  a system of   ultracold  fermions loaded onto an optical lattice made of two shifted, spin dependent sublattices. This structure  makes the CH  naturally stronger and/or comparable to normal hopping. A suitable periodic tilting of the lattice, effectively weakens the  interactions between fermions and allows to access the dynamics due to the strong CH. We describe the ensuing physics analytically  for a special value of the CH and numerically in the general case. We predict numerous  exotic properties of the system: exact and  asymmetric spin-charge separation, a gapless spectrum of lowest energy excitations, a spin-gap associated with  collective hole pairing, non-vanishing string correlator and transition to a simple ferromagnetic state.

The standard, nearest-neighbour CH  originates from inter-particle interactions,  as does the diagonal on-site interaction \cite{HubbardI}. Such CH was intensively studied for strongly correlated electrons, in particular, as a mechanism for unconventional (non-phonon mediated) superconductivity \cite{micnas,Hirsch1}, especially at weak to moderate values of the CH integral $X$. Large values of $X$ comparable to normal hopping $J$  block certain tunnelling processes resembling   effects of large repulsive Hubbard interaction \cite{Harris,Grzybowski12}.
For extreme correlations $X=J$, fermion motion is kinetically constrained so that the on-site  interaction term $U$ commutes with the rest of the Hubbard Hamiltonian. This additional symmetry facilitates an exact solution in 1-d  
\cite{Ovchinnikov,Arrachea,Korepin,Schadschneider},  revealing  several non-perturbative  effects such as a paramagnetic Mott transition (at half-filling) and exclusion statistics \cite{Vitoriano}. Furthermore, an incommensurate singlet superconducting phase appears in the proximity of the $X=J$ limit \cite{Montorsi1,Montorsi2}.  Separation of  lattices available for hopping of the two species of spins leads to a model with a different kind of  CH,  extended  
over three-sites. We study this model  in this Letter revealing markedly different phenomenology as compared to the standard CH Hubbard model.

\paragraph{The model --}
Consider an optical lattice setup with a red  and blue detuned sublattice offset by  half a lattice constant  (for related experiments on spin-dependent lattices, see \cite{Karski,Mandel03,Panahi11}), each loaded with only a single fermion spin species $\uparrow$ or $\downarrow$.
Equivalently,  one could load  two species of Tonks-Girardeau bosons \cite{Girardeau,Lieb-Liniger,Paredes,future}. 
 Each spin species only hops on its own sublattice and different fermions undergo contact interactions. As there is no mixing between sublattices, the spin index can be suppressed yielding the Hamiltonian:
\begin{gather}
H= \sum_{i}^{} (-J+X n_{i+1})(c_i^{\dagger}c_{i+2} + h.c) + \sum_{i}V n_i n_{i+1},
\label{model}
 \end{gather}
where odd (even) sites correspond to the blue (red) sublattice. The hopping $J>0$ connects next nearest neighbours.  The leading term stemming from the contact interaction is the nearest neighbour repulsion $V \gg J$, followed by the  CH to a next nearest neighbour $X>0$ conditioned on the intervening site occupancy (Fig.\ref{fig0}, details in Supplemental Materials -- SM). Interestingly, although in usual scenarios CH terms are subleading in comparison to normal hopping \cite{Dutta-review}, the separation of the sublattices yields a novel, viable path to a strong  CH term $X \sim J$.

Fast, periodic tilting of the optical lattice described by $H_S(t) = A \cos \omega t \sum_j j n_j $ \cite{Eckardt,Lignier07}, at frequency close to resonance with the interaction: $V = 2\omega +\delta$ with detuning $\delta$,  allows control over the full range of properties of the system. The  long-time dynamics of fermions is described by the  Hamiltonian $\widetilde{H}(t) = H + H_S (t) $ in the frame moving with the interaction and the lattice tilting: $H' = U\widetilde{H}U^{\dagger} + i\dot{U}U^{\dagger}$, with $U=\exp(i 2\omega t \sum_{j} n_j n_{j+1}  + i \int_{0}^{t} H_S(t')dt')$. As shown in the SM, choosing the tilting amplitude $A$ so that $J_0(2A/\omega)=J_2(2A/\omega)$, where $J_n(x)$ are the n-th order Bessel functions of the first kind, the effective Hamiltonian $H'$ is time-independent and  takes exactly the form of the original CH model (\ref{model}) with dressed parameters $J \rightarrow J J_0(2A/\omega), X \rightarrow  XJ_0(2A/\omega), V \rightarrow \delta$.  So, although the bare interaction is large for fixed ratio $X/J$, the driving makes relevant the  complete range of parameters of the model (\ref{model}) from $V=0$ (exact resonance at $\delta=0$) to $J/V=0$, controlled via the detuning and the dressing parameter $J_0(2A/\omega)$. We elucidate the model's properties below.

\begin{figure}[t]
\centering
\includegraphics[width=0.8\columnwidth, angle=0]{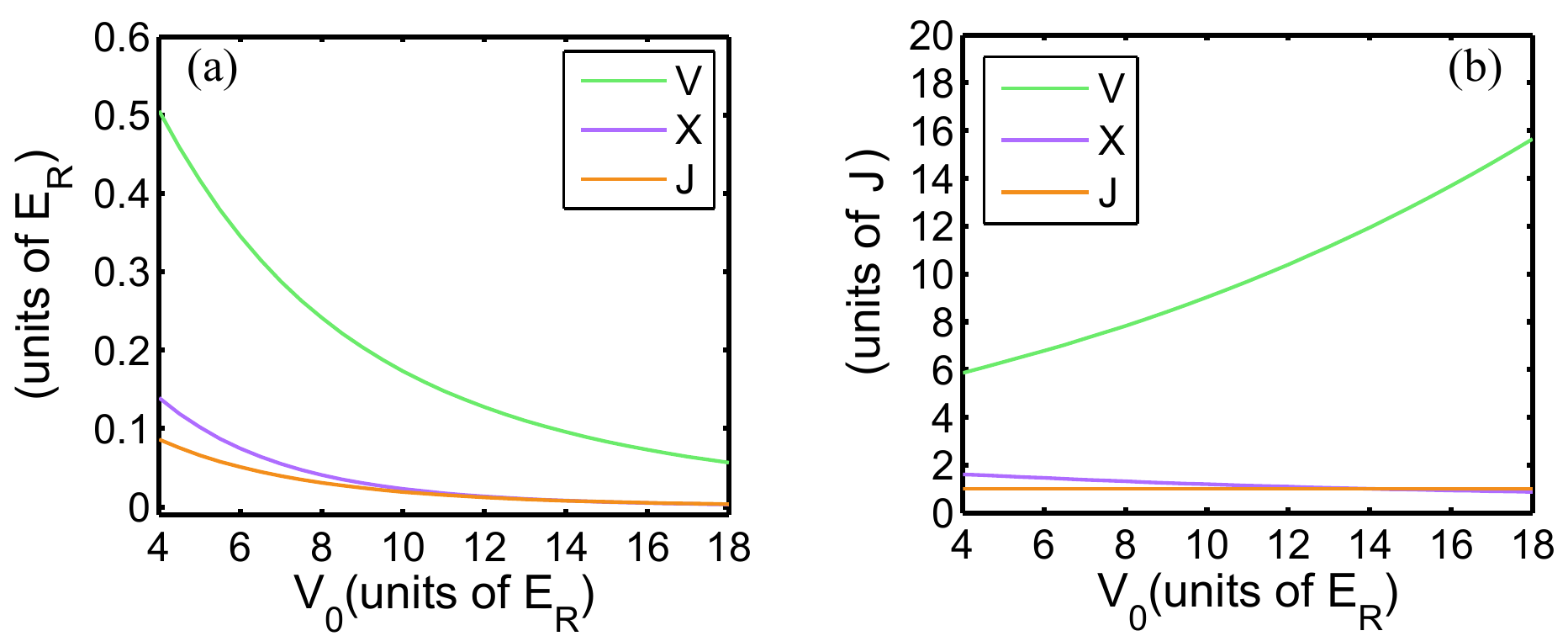}
\caption{\label{fig0}
(Color online) Model parameters vs. lattice depths $V_0$ for a scattering length $a_s=300a_0$ (Bohr radii) and harmonic trap  energy $\hbar \omega_{\rm trap}=12 E_R $ ($E_R$ -- recoil energy). For shallow lattices, $X \approx 1.6J$ and decreases to $X=J$ for a deep $V_0 \approx 14.25E_R$ attaining a value $X \approx 0.88J$ for $V_0=18E_R$.
}
\end{figure}

\paragraph{Exact solution at $X=J,V=0$ -- } 
At this point, the model has an important symmetry, as  the free and correlated hopping cancel for two nearest neighbour fermions. So, different fermions are forbidden to pass each other and  the sequence of $\uparrow$ and $\downarrow$ fermions is conserved  under open boundary conditions (OBC). The Hamiltonian can be diagonalized for each fermion sequence  separately and, as shown below, fermions may be  completely stripped of their spin. The fixed sequence in the original lattice  constrains  the space available for the motion of fermions. Thus while the original system consists of two coupled chains of length $L$, the spinless fermion degrees of freedom live on a charge-chain of length $L_{\rm eff}$,  which remarkably depends on the fermion sequence.

\begin{figure}[t]
\centering
\includegraphics[width=0.75\columnwidth, angle=0]{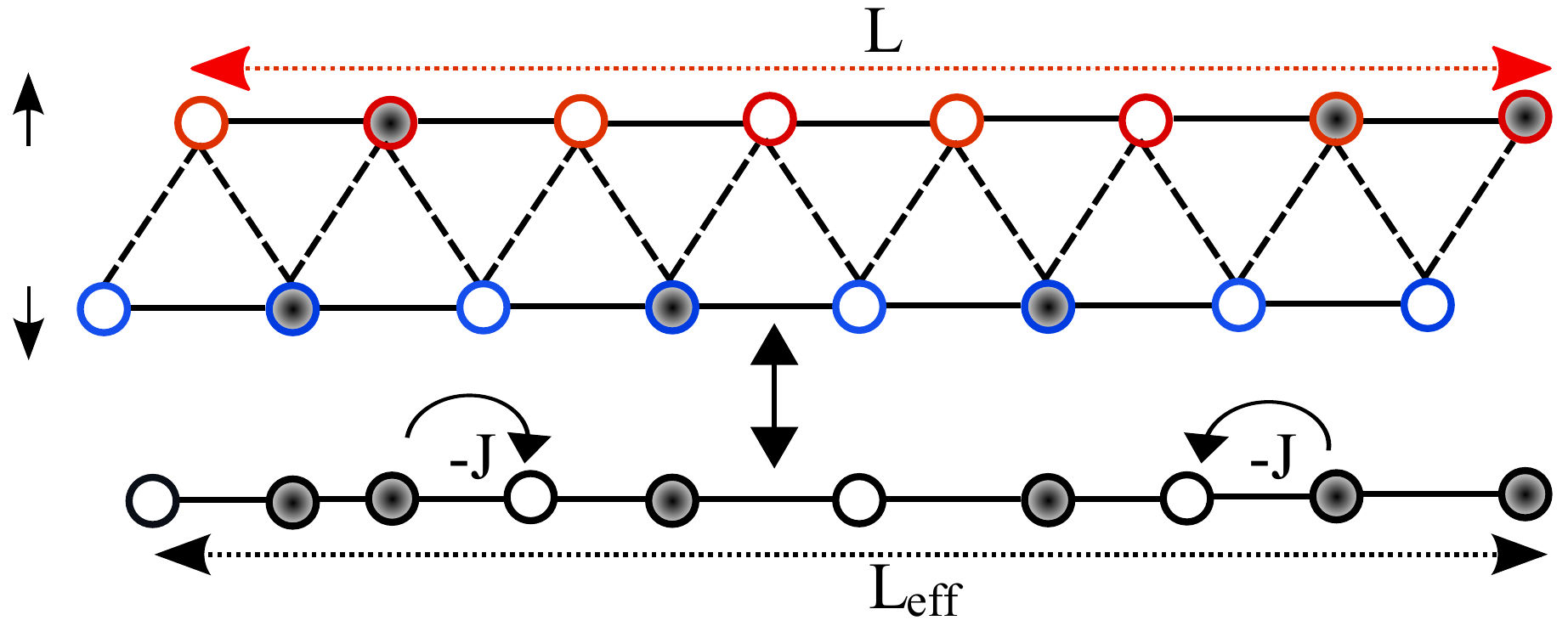}
\caption{\label{fig1}
(Color online) 
Mapping to spinless fermions in the kinetically constrained model ($X=J,V=0$) when different spins cannot pass each other. 
Occupied sites are shaded.  
Each pair of empty sites  between different spin consecutive fermions is an empty site in the charge chain. 
Empty sites on a single chain between two closest same species fermions are empty sites in the charge chain.
}
\end{figure}

 We  restrict the total number of fermions $N_t \leq L$, so as to allow the realization of all possible spin sequences. Note that the distance (number of empty sites) on the original lattice $d$ between  two consecutive fermions is always even (odd) if they belong to  different (same) species, with the minimal distance 0 (1) respectively (see Fig. \ref{fig1}). 
The distance of the first (last) fermion from the leftmost (rightmost) edge
 of the lattice is always even for $\downarrow$ ($\uparrow$) or odd for $\uparrow$ ($\downarrow$) fermion respectively. The conservation of spin order leads to the conservation of order of parities of consecutive distances. For example (see Fig.\ref{fig1}) for 6 fermions and the following spin order $\downarrow \uparrow \downarrow \downarrow \uparrow \uparrow$ 
there are 7 distances and their parity order is 
EEEOEOE.
 So, to specify any configuration, for a given spin order, one may use a set of  effective distances $d_{\rm eff}$, where $d_{\rm eff}=d/2$ or $d_{\rm eff}=(d-1)/2$ for even or odd $d$ respectively. 
The $d_{\rm eff}$ are the number of empty sites in the charge lattice. Note that the sum of these effective distances $D_{\rm eff}$ is given by $D_{\rm eff}=(D_0-M_O)/2$ where $D_0=2L-N_t$ is the sum of the distances on the real lattice and $M_O$ is the number of odd distances. This  means that $D_{\rm eff}$ is a constant for a given spin order, and there is a \textit{1-to-1 mapping} between configurations on the real lattice (for that order) and configurations of $N_t$ spinless fermions on the effective lattice of length $L_{\rm eff}=D_{\rm eff}+N_t=L+(N_t-M_O)/2$. Futhermore note that the number $S=(N_t-M_O)/2$ is the number of switches, {\it i.e.} the number of times a $\downarrow$ fermion is followed by a $\uparrow$ fermion.  Hence $L\leq L_{\rm eff}=L+S \leq L+N_\uparrow$, where $N_\uparrow$ is the number of fermions on the upper chain.

The action of $H$ (with $V=0$) in the original lattice is equivalent to the action of nearest neighbour free-fermion hopping in the charge lattice, which immediately allows the identification of all eigenstates and energies of the system. In particular the lowest energy in a given spin order sector, in the thermodynamic limit, is
\begin{gather}
 E_0/L = -\frac{2J}{\pi} \frac{L_{\rm eff}}{L} \sin \frac{\pi 2 \rho L}{L_{\rm eff}}
\label{en:1}
\end{gather}
where $\rho=N_t/(2 L)$ is the original fermion density. 
The most interesting features  of the solution are listed below:

(i) The decoupling of the spin order from the charge degrees of freedom is an intricate manifestation of  spin-charge separation ubiquitous in 1-dimensional settings \cite{Giamarchi-book}.  The novelty here is that it is asymmetric: while charge excitations can be created without affecting the spin configuration, a change in the spin order explicitly affects (via $L_{\rm eff}$ ) the charge energy spectrum.

(ii) We restrict to an even number of fermions $N_t=2N$. The global ground state corresponds to spin order of the form $|\downarrow \uparrow \downarrow \uparrow \ldots \downarrow \uparrow \rangle$. This sequence maximizes $L_{\rm eff}$ for a given $L$ and $2N$ as it corresponds to the lowest possible particle density on the charge lattice. 
Hence the spin part of the ground state wave function has perfect antiferromagnetic order. Note that this order is not degenerate with respect to flipping all spins -- this is a manifestation of  the lack of reflection symmetry of the zigzag chain about a horizontal line through the middle of the chain. 

(iii) Antiferromagnetic order in the spin sector implies the existence of hidden non-local string order in the full lattice of moving particles \cite{Kruis}, where each $\downarrow$ fermion is followed by an $\uparrow$, with an arbitrary number of holes between each pair. This is an  analog of string order in spin-1 Heisenberg chains \cite{AKLT}.  We stress that there is no local symmetry breaking of the density wave type accompanying the formation of this string order -- the ground state is translationally invariant (modulo OBC).  Unlike for spin-1 chains, the low lying excitations are gapless (in the thermodynamic limit)  corresponding to charge excitations above the Fermi sea at fixed spin order.

(iv) Although the spectrum is gapless in each order sector there is a spin gap,  i.e. the energy cost asociated with flipping one spin  which shortens the charge lattice size by one site  $L_{\rm eff}=L+N-1$.
All single particle energies are shifted  due to the change in $L_{\rm eff}$ leading, in the thermodynamic limit, to a non-zero spin gap for arbitrary filling densities.
\begin{gather}
\Delta_S =E_0 (L_{\rm eff} = L + N -1)- E_0 (L_{\rm eff} =L+N)\nonumber \\ 
\approx
\frac{2J \left((\rho +1) \sin \left(\frac{2 \pi  \rho }{\rho +1}\right)-2 \pi  \rho  \cos \left(\frac{2 \pi  \rho }{\rho +1}\right)\right)}{\pi  (\rho +1)}.
\label{spingap}
\end{gather} 
For $X=J$ switching the order of a pair of nearest neighbour fermions in the ground state entails the same energy cost.
There are $N$ such degenerate spin order configurations. Interestingly, the alternative  perfect antiferromagnetic configuration $| \uparrow \downarrow \uparrow \downarrow \ldots \uparrow \downarrow \rangle $ obtained by flipping all spins in the ground state  is also associated with a $(L+N-1)$-site charge lattice and so is exactly degenerate with the lowest lying spin excitation. 
The $\Delta_S \neq 0$ 
 suggests a protecting mechanism for the string ordered phase in our system when  moving away from the integrable case $X=J$, $V=0$.

\begin{figure}[t]
\centering
\includegraphics[width=0.7\columnwidth, angle=0]{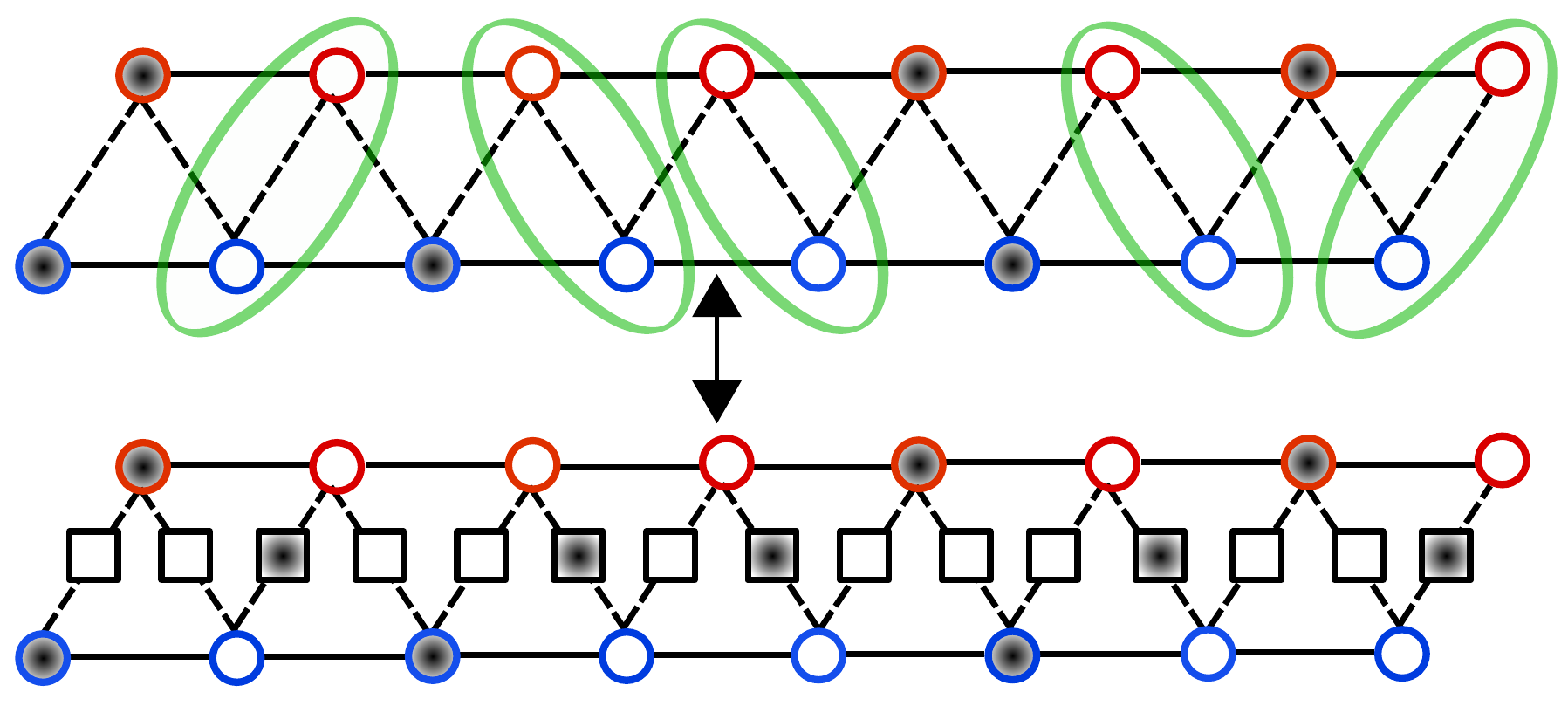}
\caption{\label{fig2}
(Color online)  All nearest  neighbour holes are uniquely paired in the ideally string ordered configurations of fermions.  
So, the motion of fermions  in the real lattice can be equivalently described as hole-pair hopping. (Below) Hole-pairs are single particles that hop on a dual lattice defined on the edges of the zig-zag chain (linear chain of squares). Shaded squares are bound hole-pairs. Uniqueness of pairing of holes (above) implies exclusion of simultaneous occupation of nearest neighbour sites on the dual lattice, i.e. infinite repulsion of paired holes on these sites. 
}
\end{figure}

\begin{figure*}[t]
\centering
\includegraphics[width=1.6\columnwidth, angle=0]{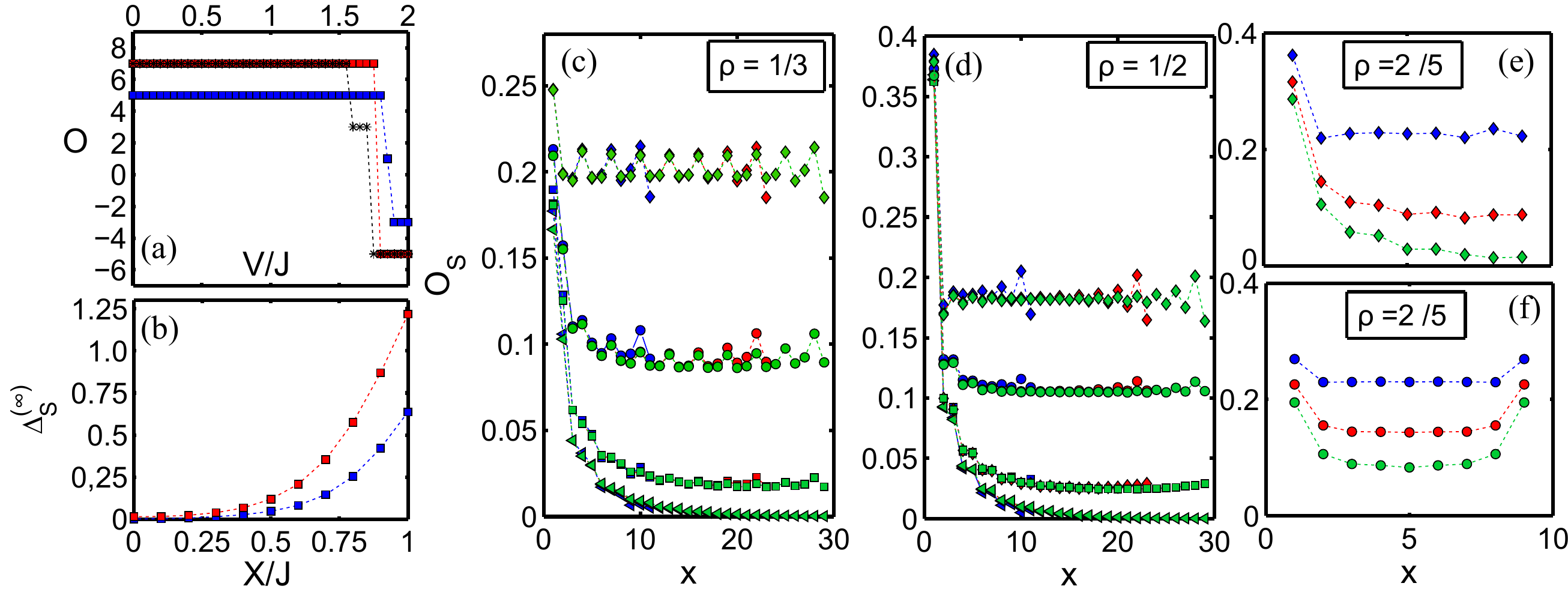}
\caption{\label{fig3}
(Color online) Exact Diagonalization (ED) (a,b,e,f) and  TEBD algorithm \cite{Vidal-TEBD,tebd} (c,d) results. (a) $O$ vs $V$ for $X=J$ where  $2N=8,2L=20$ (black), $2N=8$, $ 2L=24$ (red),$2N=6$, $2L=24$ (blue). (b) Extrapolated spin gap $\Delta_S^{(\infty)}$ to thermodynamic limit for $\rho=1/2$ (red) and $\rho=1/3$ (blue) for $V=0$. (c,d) Finite size scaling $O_S$ vs distance $x$ under OBC for different densities and $V=0$. Data are  for $X= 1.0J$ (diamonds), $ 0.6J$ (circles), $0.3J$, (squares), $0.0J$ (triangles). Compared are lattice lengths $2L = 24$ (blue), $2L=48$ (red) and $2L=60$ (green) for every $X$. (e-f) $O_S$ for on a lattice of length $2L=20$ for OBC (e) and periodic boundary conditions (PBC) (f) for $V=0$ and $X=J$, $X=0.5J$, $X=0J$ from top to bottom.  Oscillations in data in (c-e) are due to Friedel oscillations and OBC. These are diminished under PBC (f). 
}
\end{figure*}

(v) A revealing aspect of the ground state is that {\it all} holes 
are unambigiuosly bound into nearest neighbour pairs (Fig.\ref{fig2}).
The  particle motion  corresponds to the coherent motion of paired-holes through the lattice  (in the opposite direction to a particle). 
The finite spin gap $\Delta_S$ for {\it all densities}, which in the paired hole picture is the energy of breaking  paired holes, suggests the ground state  can be interpreted as an exotic hole superconductor.  Since, the system is 1-dimensional, there is no explicit symmetry breaking but rather algebraically decaying superconducting correlations (SC). Explicitly, notice that the motion of paired holes in configurations as in Fig.\ref{fig2} can be described by a 
spin-1/2 XXZ model at fixed magnetization $m=\rho/2$,
defined on the edges of the zigzag chain: $H =  \sum_{i} J(\sigma_i^{\dagger}\sigma_{i+1} +h.c) + \Delta/4 (\sigma^z_{i}+1)(\sigma^z_{i+1}+1)$, where $\sigma_z=1(-1)$ corresponds to the occupation (lack) of a paired-hole and $\Delta=\infty$. 
The ground state is a Luttinger liquid and the SC in the real lattice  correspond to the transverse spin correlation function in the dual one. The latter can be obtained using bosonization techniques (see Ch. 6 and Eq. (6.47) of Ref. \cite{Giamarchi-book}):
\begin{gather*}
\langle c_{x+1}c_{x}c^\dagger_0c^\dagger_1\rangle_{\rm real}=\langle S^+(x)S^-(0)\rangle_{\rm dual} \nonumber\\
=C\cos(2\pi m x)\Bigl(\frac{1}{x}\Bigr)^{2K+1/(2K)}+C'\cos(\pi x)\Bigl(\frac{1}{x}\Bigr)^{1/(2K)},
\label{Corrs}
\end{gather*}
where the  Luttinger parameter $K$ increases from $1/4$ for $\rho \rightarrow 0$ to $1$ for $\rho \rightarrow 1$ ~\cite{future}.

Finally, note that while the configurational mapping to the effective chain  holds for $X=J,V\neq 0$, the solely interspecies interaction negates the mapping to a spinless fermion Hamiltonian in general. However, two  spin orders are exceptions: the perfect antiferromagnetic order maps onto a $t-V$ spinless fermion model  which as seen below is stable upto finite $V$, and the fully polarized spin state which stabilizes at large interactions. In particular, for $V=\infty$ ($J/V=0$), the largest possible obtainable length is  $L_{\rm eff}=L$  yielding a ferromagnetic (FM) phase separated ground state for equal number of $\uparrow$ and $\downarrow$ spins:  $|\downarrow \downarrow \ldots \downarrow \uparrow \uparrow \ldots \uparrow\rangle$.

\paragraph{Robustness of string ordered phase -- } 
Intuitively, to induce a change in the described ideal string configuration for $X=J$, the energy penalty associated with $V \neq 0$ should be large enough so as to offset the kinetic energy gained via the introduction of defects in the spin order. In order to analyse this, we introduce the parameter:
\begin{gather}
 O = \sum_{j=1}^{2L} \sum_{j'>j} n_j Q_{j,j'} n_{j'}e^{i\pi(j-j'+1)},
\label{AverageJumps}
\end{gather} 
where $Q_{j,j'}=\prod_{j'>l>j}(1-n_l) $  for $j'>j+1$ or  $Q_{j,j'}=1$ where $j,j'$ are subsequent sites.
 With an equal number $N$ of $ \uparrow  $ and $\downarrow$ fermions,   $O=2N-1$ is maximal for ideal string ordered configurations, and minimal for the phase separated state with $O=3-2N$. 
Ground state numerical results (see Fig.\ref{fig3}(a)) show that for arbitrary densities, a large region of stability of ideal string order (and thus collective hole pairing) is followed by a first order transition to the FM state for $V< 2J$. 
In some cases, the jump - in these finite size systems - is via an intermediate spin sequence. It is interesting  that the simple variant of itinerant ferromagnetism  unambiguously exists here for realistic  parameters. In particular, without lattice driving and fine tuning, the bare interaction (Fig.\ref{fig0})  places the system in the FM phase.

For $X\neq J$, the spin sequence is subject to dynamical changes.
We restrict to $X/J \leq 1$ due to the symmetry $X \rightarrow 2J-X$  of the model \eqref{model} \cite{future}.  
Without CH  ($X=0$, $V \neq 0$) an interaction driven, direct  liquid to phase-separated jump transition  for arbitrary fillings already exists \cite{Yarlagadda}.  So, we shall here not consider the similar expected effect  of $V\neq 0$ but rather the fate of the string ordered state for $V=0$. First, note that the spin-gap $\Delta_S$ is always open for $X>0$ (Fig. \ref{fig3}(b)) indicating hole pairing and  long range string correlations for arbitrary low values of  CH. 
  To probe long range non-local hidden  order for general $X\neq J$, we introduce a den Nijs-Rommelse type string parameter \cite{Nijs,Kruis}, $O_S$. Grouping pairs of nearest  neigbour sites into dimers $(1,2), (3,4), \ldots (2L-1,2L)$, define the  operator $S^z_{ i} = (n_{2i} - n_{2 i -1})$ with values $+1,0,-1$ in analogy with a spin-1 $S_z$ operator, so that
\begin{gather}
O_S(m,m')= \langle S^z_{m} \exp\Big(i \pi \sum_{m<l<m'} S^z_{l}\Big)S^z_{m'} \rangle.
\label{StringOrder}
\end{gather} 
Figs. \ref{fig3}(c-f), clearly show that $O_S$ attains a finite value for the ideally string ordered ground state at $X=J$. Reducing $X$ leads to a diminishing, yet finite large distance saturation value of $O_S$ for $X\neq 0$ with no evidence of destruction of  long range string order before $X=0$.
We have checked that this holds for various densities $\rho$.

\paragraph{Final comments -- }  Long range string order is often associated with a symmetry protected topological phase characterized by existence of edge modes and topological degeneracy. Our model, under OBC, breaks inversion symmetry which has been shown recently to protect the existence of such features \cite{Oshikawa}.  Indeed, the string ordered ground state is unique in the thermodynamic limit, since the spin gap separates it from the configuration with all spins flipped. This feature is distinct from the standard Haldane phase --  there are no zero energy edge modes yielding ground state degeneracy.

While string order parameters were first shown to characterize gapped spin-1 chains \cite{AKLT,Nijs}, recent studies have shown that  specific non-local order parameters naturally describe a plethora of gapped phases without local order in 1-d systems. However, these commonly  take the form of parity correlators that  capture the confinement of  quantum fluctuations {\it e.g.} in Mott insulator phases of bosons \cite{Endres} and fermions as well as the Luther-Emery phases of the attractive Hubbard model \cite{Montorsi3,Montorsi4}. Intriguingly, however, critical string order (algebraically decaying) was shown to underlie spin-charge separation in 1-d systems  surviving upto the limit of non-interacting fermions \cite{Kruis}. It is interesting that the asymmetric type of spin-charge separation encountered in the model considered in this Letter leads to a truly long range string order. 
Moreover, the lowest lying charge excitations in the system are gapless and our model thus constitutes a rare example of finite string order in a gapless phase (see also Refs. \cite{Altman2, Sarma}) of itinerant particles.

\paragraph{Acknowledgements --}
We thank Roman Micnas for comments and discussions. 
We acknowledge support from EU grants OSYRIS (ERC-2013-AdG Grant No. 339106), SIQS
(FP7-ICT-2011-9 No. 600645), QUIC (H2020-FETPROACT-
2014 No. 641122), EQuaM (FP7/2007-2013 Grant No.
323714), Spanish Ministry grant FOQUS (FIS2013-46768-P),
as well as the Polish National Science Center Grants No.
DEC-2011/03/B/ST2/01903 and DEC-2012/04/A/ST2/00088. 
 R.W.C. acknowledges a Mobility Plus ´
fellowship from the Polish Ministry of Science and Higher
Education.

\section{Supplemental Materials:}

\section{Effective parameters of the Correlated Hopping model}

Our system consists of two-species of ultracold atoms trapped in a 1-d optical lattice. For one species, we consider a blue detuned laser and for the other a red-detuned laser to create the optical lattice. As a result, the respective lattices for the species are shifted in space by a half-lattice constant $a/2$. The lattice potentials are therefore $V^\downarrow_{\rm latt}(x)=V_\downarrow \cos^2\pi x/a$ and $V^\uparrow_{\rm latt}=V_\uparrow \cos^2(\pi x/a+\phi)$ where the red and blue-detuned case corresponds to $\phi=\pi/2$. The corresponding Hamiltonians reads,
\begin{widetext}
\begin{gather}\label{hamtotal}
H = \int dx \sum_{\sigma=\uparrow,\downarrow}\Psi^{\dagger}_\sigma(x)
\left [- \frac{\hbar^2}{2m_\sigma}\frac{d^2}{dx^2} +  V^{\sigma}_{\rm latt}(x) \right]\bm{\Psi}_\sigma(x)
+ g_{\rm 1D}\int dx \bm{\Psi}^{\dagger}_a(x)\bm{\Psi}^{\dagger}_b(x)\bm{\Psi}_b(x)\bm{\Psi}_a(x), 
\tag{S1}%
\end{gather}
\end{widetext}
where $\bm{\Psi}^{\dagger}_\sigma(x),\bm{\Psi}_\sigma(x)$ are the creation and annihilation field operators for the species $\sigma=\uparrow,\downarrow$. 
The first term in $[..]$ describes the single particle kinetic and potential energy while the second term describes the one-dimensional contact interaction with strength 
$
g_{\rm 1D}=\frac{4\pi\hbar^2}{m_r} \frac{a_{\rm s}}{\pi a^2_{\rm trap}} \frac{1}{1-C a_{\rm s}/a_{\rm trap}},
$
with reduced mass $m_r=2m_\uparrow m_\downarrow /(m_\uparrow+m_\downarrow)$, scattering length $a_{\rm s}$ and the transverse oscillator length $a_{\rm trap}$ and constant $C \approx 1.46$ \cite{Olshanii98}.
Next we expand the field operators, $\bm{\Psi}^{\dagger}_\sigma(x)=\sum_{i_{\sigma}}\omega_{i_\sigma}(x)\bm{\sigma}^{\dagger}_{i_\sigma}$, where $i_{\sigma}$ denotes the sites for the species $\sigma=\uparrow,\downarrow$, $\omega_{i_\sigma}(x)$ are the respective Wannier functions and $\bm{\sigma}^{\dagger}_{i_\sigma}$ are the creation operator for the species $\sigma$ at site $i_{\sigma}$. The contact interaction leads to both nearest neighbour repulsion and correlated hopping in the Wannier basis. Indeed, the corresponding tight-binding Hamltonian in terms of the local operators reads:
\begin{widetext}
\begin{gather}\label{hamlocal}
H = - \sum_{\sigma}\sum_{i_\sigma}J_{\sigma}(\bm{\sigma}^{\dagger}_{i_\sigma}\bm{\sigma}_{i_\sigma+1}+h.c) + X \sum_{i_\uparrow,i_\downarrow}(\bm{\sigma}^{\dagger}_{i_\uparrow}\bm{n}^\sigma_{i_\downarrow}\bm{\sigma}_{i_\uparrow+1}+\bm{\sigma}^{\dagger}_{i_\downarrow}\bm{n}^\sigma_{i_\uparrow+1}\bm{\sigma}_{i_\downarrow+1}+h.c) + V \sum_{i_\uparrow,i_\downarrow} (\bm{n}^a_{i_\uparrow}+\bm{n}^a_{i_\uparrow+1})\bm{n}^b_{i_\downarrow}.
\tag{S2}
\end{gather}
\end{widetext}
To express the parameters, first we assume that $m_\uparrow=m_\downarrow=m$ and the lattice depths are same, $V_\uparrow=V_\downarrow=V_0$. Moreover, we express the lengths $x\rightarrow x/a$ and the energy with respect to recoil energy $E_R=\pi^2\hbar^2/2ma^2$. This implies that $J_\uparrow=J_\downarrow=J > 0$ and for Wannier functions we use the convention, $\omega_{i_\uparrow}(x)=\omega(x-x_i)$ and $\omega_{i_\downarrow}(x)=\omega(x-x_i-1/2)$. 
With this convention  Eq.(S2) is equivalent to the concisely written model Eq.(1) in the main text with suppressed spin indices. 
The parameters are now expressed as,
\begin{gather}
X = g\int dx \omega(x) \omega(x-1) \omega^2(x-1/2), \nonumber\\ 
V = g\int dx \omega^2(x) \omega^2(x-1/2)
\tag{S3}
\end{gather}
with
$
g = \frac{8}{\pi}\frac{a_{\rm s}a}{\pi a^2_{\rm trap}} \frac{1}{1-C a_{\rm s}/a_{\rm trap}}
$. The trapping potential in Fig.1 of the main text corresponds to $a_{\rm trap}/a \approx 0.14$.

\section{Periodic lattice modulation and effective time independent Hamiltonian}

We consider the system described in the previous section to be subject to shaking of the optical lattices with driving frequency $\omega$ and tilting amplitude $A$. The time-dependent Hamiltonian is then given by:
\begin{gather}
\widetilde{H}(t) =  H + A\cos(\omega t) \sum_j j n_j
\label{shaken}
\tag{S4}
\end{gather}
where we represent the Hamiltonians in concise form corresponding to Eq.(1), {\it i.e.} $j$ enumerates consecutive sites on the zigzag chain.

The interaction strength $V$ is the largest energy scale in the system. 
We choose the driving frequency to satisfy the quasi-resonance condition  $V = 2 \omega + \delta$, where $\delta$ is the (small) detuning. In order to eliminate the fast degrees of freedom in the problem, we transform to the co-rotating frame  
via the unitary transformation $U = \exp(i 2\omega t \sum_{j} n_j n_{j+1}  + i A/\omega \sin \omega t \sum_j jn_j)$ to obtain the effective Hamiltonian $H' = UHU^{\dagger} + i \dot{U} U^{\dagger}$:
\begin{gather}
H' = \sum_i {\cal G}_i(t) (-J + X n_i)  c_i^{\dagger} c_{i+2} + h.c. + \delta \sum_i n_i n_{i+1}
\label{effective}
\tag{S5}
\end{gather}
where the hopping and CH terms are modulated by the time dependent function
\begin{gather}
{\cal G}_i(t) = \exp \Bigg( - i 2  \omega t \Big(n_{i+3} - n_{i-1} \Big) - i \frac{2A}{\omega} \sin\omega t\Bigg)
\label{modulation}
\tag{S6}
\end{gather}
The occupation number operator is a projector $n^2=n$, so  
$\exp(i Bt n) = (1-n) + n\exp(i B t)$,
and therefore
\begin{widetext}
\begin{gather}
\exp(-i Bt (n_1-n_2)) = (1-n_1-n_2+2n_1n_2) + n_1\exp(-i Bt) + n_2\exp(iBt) -n_1n_2(\exp(iBt) + \exp(-iBt)).
\label{expansion}
\tag{S7}
\end{gather}
\end{widetext}
Using this and the Jacobi-Anger identity,
\begin{gather}
\exp\Big(i \frac{2A}{\omega}\sin \omega t\Big) = \sum_{m=-\infty}^{+\infty} J_{m}\Big(\frac{2A}{\omega}\Big) \exp(i m \omega t)
\tag{S8}
\end{gather}
we obtain the following time independent effective modulation  ${\cal G}$ on averaging out the fast oscillating terms:
\begin{gather}
{\cal G}_i =  (1-n_{i+3}-n_{i-1}+2n_{i+3}n_{i-1}) J_{0}\Big(\frac{2A}{\omega}\Big) \nonumber\\
+ (n_{i+3}+n_{i-1}-2n_{i+3}n_{i-1}) J_{2}\Big(\frac{2A}{\omega}\Big),
\label{effective:1}
\tag{S9}
\end{gather}
after applying the relation $J_{2}(x) = J_{-2}(x)$ for the Bessel functions of the first kind.
The time averaging procedure can be carried out as long as $\omega \gg J,X,\delta$ which is the case here (see Fig.1 in main text).
In general, the  lattice  modulation generates CH terms, through ${\cal G}_i $, beyond those considered in the paper. However, choosing the tilting amplitude $A^{*}$ such that $J_0 (2A^{*}/\omega) = J_{2}(2A^{*}/\omega)$,  eliminates these completely and one obtains the CH model Eq.(1) with the bare hopping and CH terms renormalized by ${\cal G}=J_0(2A^{*}/\omega)$, {\it i.e.} 
\begin{gather}
H' = \sum_i   J_{0}\Big(\frac{2A^{*}}{\omega}\Big) (-J + X n_i)  c_i^{\dagger} c_{i+2} + h.c. + \delta \sum_i n_i n_{i+1}
\label{effective}
\tag{S10}
\end{gather}
The hopping  renormalization is given by $J_0(2A^{*}/\omega \approx 1.84118) = 0.316$ at the first intersection point of the Bessel functions.  The residual nearest neighbour interaction now stems solely from the detuning $\delta$ and in particular can be chosen to be equal to zero as well as non-zero values - both cases are considered in the main text. Notice that choosing larger amplitudes $A^*$ corresponding to subsequent intersection points leads to a fast diminishing of the effective normal and correlated hoppings thus allowing to access smaller  effective hopping to interaction ratios.


\begin{thebibliography}{100}


\bibitem{Giamarchi-book} T. Giamarchi,  \textit{Quantum Physics in One Dimension} (The International Series of Monographs on Physics),  Clarendon Press, first edition (2004).


\bibitem{Tomonaga} S. Tomonaga, Prog. Theor. Phys. \textbf{5}, 544 (1950).

\bibitem{Luttinger} J. M. Luttinger, J. Math. Phys. \textbf{4}, 1154 (1963). 

\bibitem{Haldane} F. D. M. Haldane, J. Phys. C: Solid State Phys. \textbf{14}, 2585 (1981).


\bibitem{Ogata} M. Ogata and H. Shiba, Phys. Rev. B \textbf{41}, 2326 (1990).

\bibitem{Faddeev} L. D. Faddeev and L. A. Takhtajan, Phys. Lett. A \textbf{85}, 375 (1981).





\bibitem{AKLT} I. Affleck, T. Kennedy, E. H. Lieb, and H. Tasaki, Phys. Rev. Lett. {\bf 59}, 799 (1987); I. Affleck, T. Kennedy, E. H. Lieb, and H. Tasaki, Commun. Math. Phys. {\bf 115}, 477 (1988).

\bibitem{Nijs} M. den Nijs and K. Rommelse, Phys. Rev. B {\bf 40}, 4709 (1989).

\bibitem{Dutta-review} O. Dutta, M. Gajda, P. Hauke, M. Lewenstein, D.-S. L{\"u}hmann, B. A. Malomed, T. Sowi{\'n}ski, J. Zakrzewski, Rep. Prog. Phys. \textbf{78},  066001 (2015).


\bibitem{Jurgensen} O. J{\"u}rgensen, F. Meinert, M. J. Mark, H.-C. N{\"a}gerl, D.-S. L{\"u}hmann,  Phys. Rev. Lett. \textbf{113}, 193003 (2014); D.-S. L{\"u}hmann, O. J{\"u}rgensen, and K. Sengstock, New Journal of Physics 14, 033021 (2012).

\bibitem{Sowinski}  T. Sowi{\'n}ski, O. Dutta, P. Hauke, L. Tagliacozzo and M. Lewenstein, Phys. Rev. Lett. \textbf{108}, 115301 (2012).

\bibitem{Santos} \'A. Rapp, X Deng,  and L. Santos, Phys. Rev. Lett. {\bf 109},  203005 (2012).

\bibitem{HubbardI} J. Hubbard, Proc. Roy. Soc. London {\bf 276}, 238 (1963).

\bibitem{micnas} R. Micnas, J. Ranninger, S. Robaszkiewicz, Phys. Rev. B {\bf 39}, 11653 (1989). 

\bibitem{Hirsch1} F. Marsiglio  and J. E. Hirsch, Phys. Rev. B {\bf 41}, 6435 (1990).




\bibitem{Harris} A. B. Harris and R. V. Lange, Phys. Rev. 157, 295 (1967).
\bibitem{Grzybowski12} P. R. Grzybowski and R. W. Chhajlany, Phys. Stat. Sol. \textbf{B} 249, 2231 (2012).

\bibitem{Ovchinnikov} A. A. Ovchinnikov, Mod. Phys. Lett. B {\bf 7}, 1397 (1993); A. A. Ovchinnikov, J. Phys. Cond. Mat. 6,  (1994).


\bibitem{Arrachea} L. Arrachea and A. A. Aligia, Phys. Rev. Lett. \textbf{73}, 2240 (1994). 
\bibitem{Korepin} J. de Boer, V. E. Korepin, and A. Schadschneider, Phys. Rev. Lett. \textbf{74}, 789 (1995).
\bibitem{Schadschneider} A. Schadschneider, Phys. Rev. B {\bf 51}, 10386 (1995).
\bibitem{Vitoriano} C. Vitoriano and M. D. Coutinho-Filho, Phys. Rev. Lett. {\bf 102}, 146404 (2009).


\bibitem{Montorsi1} A. A. Aligia, A. Anfossi, L. Arrachea, C. Degli Esposti Boschi, A. O. Dobry, C. Gazza, A. Montorsi, F. Ortolani, and M. E. Torio, Phys. Rev. Lett. \textbf{99}, 206401  (2007).

\bibitem{Montorsi2} A. Anfossi, C. D. E. Boschi, A. Montorsi  and F. Ortolani,
Phys. Rev. B \textbf{73}, 085113 (2006).




\bibitem{Karski} M. Karski, L. F{\"o}rster, J.-M. Choi, A. Steffen, W. Alt, D. Meschede and A. Widera, Science \textbf{325}, 174 (2009);
L. F{\"o}rster, M. Karski, J.-M. Choi, A. Steffen, W. Alt, D. Meschede, A. Widera, E. Montano, J. H. Lee, W. Rakreungdet, and Poul S. Jessen, Phys. Rev. Lett. \textbf{103}, 233001 (2009).

\bibitem{Mandel03} O. Mandel, M. Greiner, A. Widera, T. Rom, T. W. H{\"a}nsch, and I. Bloch, Phys. Rev. Lett. \textbf{91}, 010407 (2003).
O. Mandel, M. Greiner, A. Widera, T. Rom, T. W. Hansch, and ¨
I. Bloch, Nature \textbf{425}, 937 (2003).

\bibitem{Panahi11} P. Soltan-Panahi, J. Struck, P. Hauke, A. Bick, W. Plenkers, G. Meineke, C. Becker, P. Windpassinger, M. Lewenstein, and Klaus Sengstock, Nature Physics \textbf{7}, 434--440 (2011).

\bibitem{Girardeau} M. Girardeau, J. Math. Phys. \textbf{1}, 516–523 (1960).
\bibitem{Lieb-Liniger} E. H. Lieb, and W. Liniger, Phys. Rev. \textbf{130}, 1605 (1963).

\bibitem{Paredes} B.  Paredes, A. Widera, V. Murg, O. Mandel, S. F{\"o}lling, I. Cirac, G. V. Shlyapnikov, T. W. H{\"a}nsch and I. Bloch, Nature \textbf{429}, 277 (2004).

\bibitem{future} A more detailed presentation of properties of our model will be presented elsewhere in P. R. Grzybowski {\it et al.}, (in preparation).



\bibitem{Eckardt}  F. Grossmann, T. Dittrich, P. Jung, and P. Hänggi
Phys. Rev. Lett. \textbf{67}, 516 (1991); A. Eckardt, C. Weiss, and M. Holthaus, Phys. Rev. Lett. \textbf{95}, 260404 (2005).

\bibitem{Lignier07} H. Lignier, C. Sias, D. Ciampini, Y. Singh, A. Zenesini, O. Morsch, and E. Arimondo, Phys. Rev. Lett. \textbf{99}, 220403 (2007).


\bibitem{Kruis} H. V. Kruis, I. P. McCulloch, Z. Nussinov and J. Zaanen, Europhys. Lett. \textbf{65}, 512 (2004); H. V. Kruis, I. P. McCulloch, Z. Nussinov, and J. Zaanen Phys. Rev. B \textbf{70}, 075109 (2004).



\bibitem{Yarlagadda} A. Ghosh and S. Yarlagadda, Phys. Rev. {\bf B} 90, 045140 (2014). 









\bibitem{Vidal-TEBD} G. Vidal, Phys. Rev. Lett. 
\textbf{91}, 147902 (2003); {\it ibid.} \textbf{93}, 040502, (2004).
\bibitem{tebd} We have used a modified version of the Open-source TEBD implementation from http://sourceforge.net/projects/opentebd/. 







\bibitem{Oshikawa} F. Pollmann, A. M. Turner, E. Berg, and M. Oshikawa, Phys. Rev. B {\bf 81}, 064439 (2010).

\bibitem{Endres} M. Endres, M. Cheneau, T. Fukuhara, C. Weitenberg, P. Schau{\ss}, C. Gross, L. Mazza, M.C. Ba{\~n}uls, L. Pollet, I. Bloch, Science {\bf 334}, 200 (2011).



\bibitem{Montorsi3} A. Montorsi and M. Roncaglia, Phys. Rev. Lett {\bf 109}, 236404 (2012).

\bibitem{Montorsi4} L. Barbiero, A. Montorsi and M. Roncaglia, Phys. Rev. B {\bf 88}, 035109 (2013).

\bibitem{Sarma}  J. P. Kestner, B. Wang, J. D. Sau, and S. Das Sarma, Phys. Rev. B {\bf 83}, 174409 
(2011).

\bibitem{Altman2} J. Ruhman, E. G. Dalla Torre, S. D. Huber, and E. Altman, Phys. Rev. B {\bf 85}, 
125121 (2012).



\end{thebibliography}

\begin{thebibliography}{10}
\bibitem{Olshanii98} M. Olshanii, Phys. Rev. Lett. \textbf{81}, 938 (1998).
\end{thebibliography}
\end{document}